\documentclass{ifae}

\begin{document}

\begin{header}
  \title{Beyond the Standard Model: \\ Expectations at the LHC }

  \begin{Authlist}
    B.~Mele\Iref{ro}

  \Affiliation{ro}{INFN, Sezione di Roma, and 
  Universit\`a di Roma ``La Sapienza", Rome, Italy}
  \end{Authlist}

  \begin{abstract}
  In this talk, I review the main motivations for expecting new physics 
  at the TeV energy scale, that will be explorable
  at the CERN Large Hadron Collider.
 
  \end{abstract} 
  
\end{header}

\section{Introduction}
The CERN Large Hadron Collider (LHC) is expected to collide the first
proton beams in 2007. This machine has been designed to
reach c.m. $pp$ collision energies of 14 TeV and integrated luminosities 
of the order of 100 fb$^{-1}$ per year (even higher c.m. energies
will be reached in the ion-ion collision modes).
Such a proton collider is the best instrument we can conceive, 
to our present knowledge, to explore the behavior of  nature at the
energy scale of order 1 TeV or, correspondingly, at length scale of 
order $10^{-17}$cm, that is about 1/10 the ones explored so far
at past and present accelerators.

The project is extremely challenging, and costly, too.
It is requiring an incredible effort in different fields
by a very large number of people.
Then, in order to justify such an effort,
one would like to have a sort of {\it theoretical guarantee}
on its discovery potential.

The point I will try to make in my talk is about what we expect to learn
from this machine from a theoretical point of view.
The outcome will be that we may be  approaching a  {\it revolution}
in our understanding of the physics of fundamental interactions.

\section{Comparison with the CERN ppbar Collider}
In order to better assess the quality of our expectations for new physics 
at LHC, it can be useful to recall  
the theoretical expectations in 1981, 
that is just before the starting of the CERN ppbar Collider,
where proton-antiproton collisions would have been realized at the 
initial unprecedented c.m. energy of 540 GeV.
At that time, on the basis of a huge amount of experimental data, the 
Standard Model (SM) of fundamental interactions had been built up.
This was the end of a long and elaborated process whose final milestones
were
\begin{itemize}
\item 1966/67 \hskip 1cm unified description of the weak and electromagnetic
interactions by a gauge theory based 
$~~~~~~~~~~~~~~~~~~~~~~~~~~$ 
on the group $SU(2)\times U(1)$ by Weinberg and Salam ;

\item 1971 \hskip 1.5cm proof of the renormalizability of the theory  by 
t'Hooft and Veltman ;
\item 1973 \hskip 1.5cm discovery of neutral currents at CERN ;
\item 1979 \hskip 1.5cm Nobel Prize to Weinberg, Salam and Glashow .
\end{itemize}
The outcome of the process was a very solid and predictive theory 
that described
{\it observed} interactions. New particles were predicted at the starting
of the ppbar Collider :
the vector bosons $W$ and $Z$, the Higgs boson and the top quark.
At the same time, in the 70's, the Quantum Chromo-Dynamic (QCD)
was developed to describe the strong interactions.
One basic prediction of the theory was the asymptotic freedom, giving rise 
to the plethora of new phenomena connected to jet physics, also to be tested at
the ppbar Collider.
Altogether the  physics expectations at the ppbar Collider
at its starting time were solid and quite well defined.

\section{Today Expectations}
What are instead the present expectations, about three years before the
starting of LHC ?
The SM got incredibly strengthened after about 25 years of 
more and more accurate
experimental tests, not only at the ppbar Collider, 
but also at LEP, TeVatron, and lower-energy experiments.
We know today that the model based on the gauge group
$SU(3)\times SU(2)\times U(1)\; $ is $\;$ {\it the} 
theory that correctly describes 
the fundamental interactions up to energy scales $Q\sim 100$ GeV
(or down to length scales $\sim 10^{-16}$cm).
There is just one missing part : 
the conclusive test of the mechanism of the Electro-Weak Symmetry Breaking 
(EWSB). The Higgs boson has not yet been observed.
On the other hand, in recent years (especially at LEP), we got crucial
experimental informations on the Higgs sector. First of all,
we know that the SM Higgs boson must be heavier than 114.4 GeV (at 95\%C.L.),
as established from direct searches at LEP through the process
$e^+e^-\to HZ$ \cite{lephiggs}. 
A crucial, and totally independent from the latter,
piece of information arises by imposing the consistency of the SM
radiative-correction pattern in the precision-measurement sector.
\begin{figure}[hbtp]
  \begin{center}
    \epsfig{file=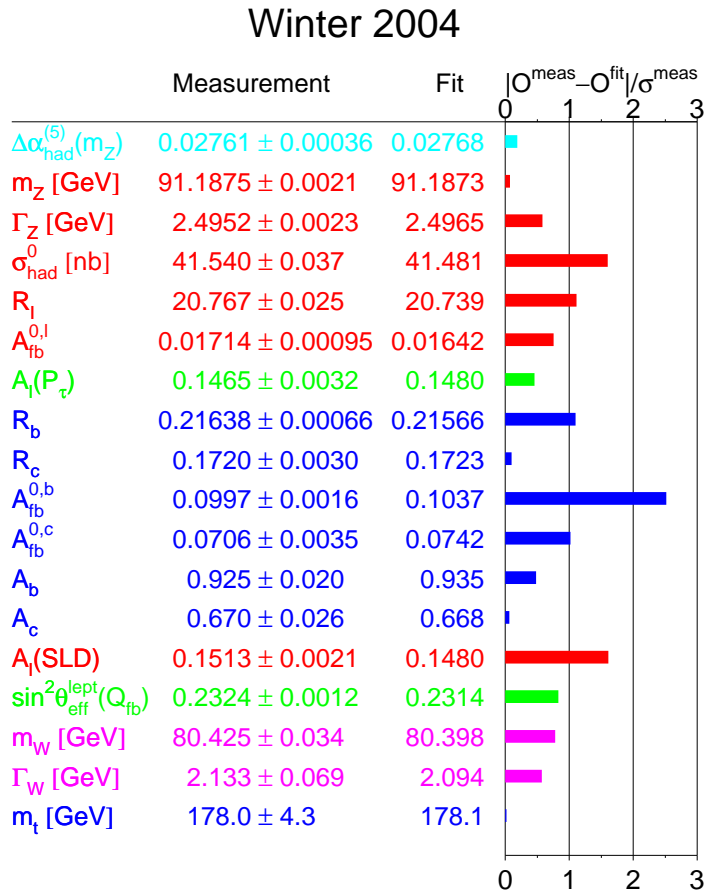,width=0.6\linewidth}
 \vskip -2.5 cm
    \caption{Summary of electroweak precision measurements \protect\cite{lepeww}.}
    \label{fig:fig1}
  \end{center}
\end{figure}
In Fig.~\ref{fig:fig1}, the present status of the fit of the SM
radiative corrections to the electroweak precision measurements at 
high $Q^2$ is shown \cite{pretest}. Apart from $Z$-pole data,
it includes TeVatron measurements of $M_W$ and the top mass, 
and the contribution to $\alpha(M_Z)$ of the hadronic vacuum polarization.
The outcome of this fit is a lower limit on $m_H$ that is milder than the direct
limit, and, most importantly, a higher limit of 237 GeV (at 95\%C.L.) 
\cite{lepeww}, that is quite tighter than the theoretical upper limit
in the minimal SM $m_H< 600-800$ GeV \cite{Hambye:1996wb}. 
The fact that the high accuracy of the electroweak measurements
(in many cases of the order of a few per mil) translates into a not-so-narrow
range for $m_H$ arises from the mild $\log m_H$ dependence of the SM radiative
pattern.

A few comments on the quality of the fit are in order.
Although in general one can state a good agreement with the SM predictions,
there are areas that manifest some tension \cite{ag}. 
\begin{figure}[hbtp]
\vskip -0.5cm
  \begin{center}
    \epsfig{file=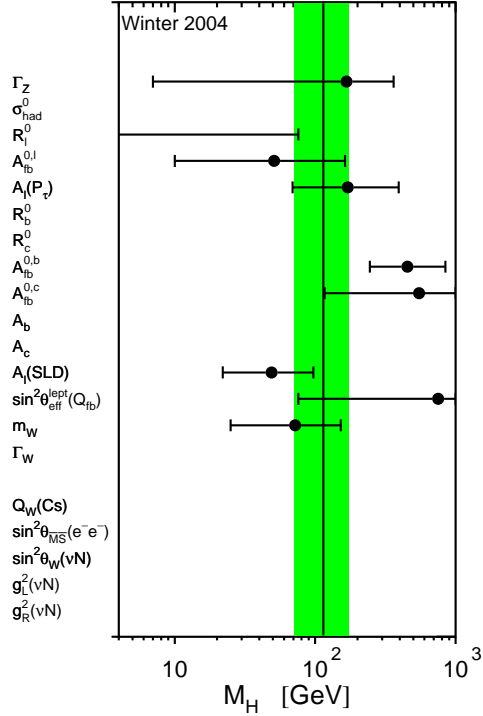,width=0.4\linewidth}
    \caption{Higgs-boson mass determination from different observables
    \cite{lepeww}.}
    \label{fig:fig2}
  \end{center}
\end{figure}
In Fig.~\ref{fig:fig2}, the $m_H$ values obtained from the most sensitive
measurements to $m_H$ is shown. One can see that both 
the leptonic asymmetries
and $m_W$ tend to favorite  $m_H$ values very close to the direct experimental
limit, while hadronic asymmetries prefer a larger $m_H$ value.
This effect is even enhanced when considering 
the two most precise $m_H$ determinations, that come from the lepton asymmetry
${\cal A}_\ell$, as measured by the left-right polarization asymmetry at SLD,
and the forward-backward asymmetry,
${\cal A}_{fb}^{0,b}=\frac{3}{4}\,{\cal A}_e{\cal A}_b$, 
measured in the $b\bar b$ production at LEP.
Here, we define as usual 
${\cal A}_f=\frac{g^2_{Lf}-g^2_{Rf}}{g^2_{Lf}+g^2_{Rf}}$,
with $g_{Lf}/g_{Rf}$  the left/right-handed $f$ fermion 
coupling to the $Z$.

A similar discrepancy can be found in the two determinations of the effective
electroweak mixing angle $\sin^2\theta^{lept}_{eff}$ coming from 
${\cal A}_\ell(SLD)$ and from ${\cal A}_{fb}^{0,b}$, 
that differ by about 3 standard
deviations.
\begin{figure}[hbtp]
  \begin{center}
    \epsfig{file=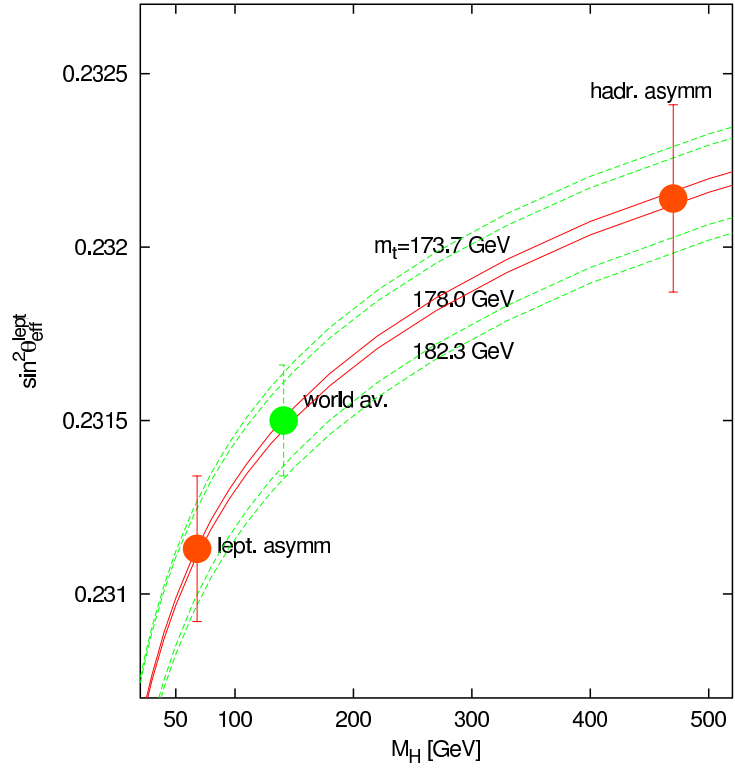,width=0.6\linewidth}
    \caption{Determinations of $\sin^2\theta^{lept}_{eff}$ from leptonic
    and hadronic asymmetries versus $m_H$ (updated from \cite{Gambino:2003xc}).}
    \label{fig:fig3}
  \end{center}
\end{figure}
In Fig.~\ref{fig:fig3}, the results for 
$\sin^2\theta^{lept}_{eff}$ as determined
starting from either all leptonic asymmetries or all hadronic asymmetries 
(and their combination) are compared versus
$m_H$, placing each result at the $m_H$ value that would correspond  
to the central value of the top mass, $m_{t}$ \cite{Gambino:2003xc}.

\begin{figure}[hbtp]
  \begin{center}
    \epsfig{file=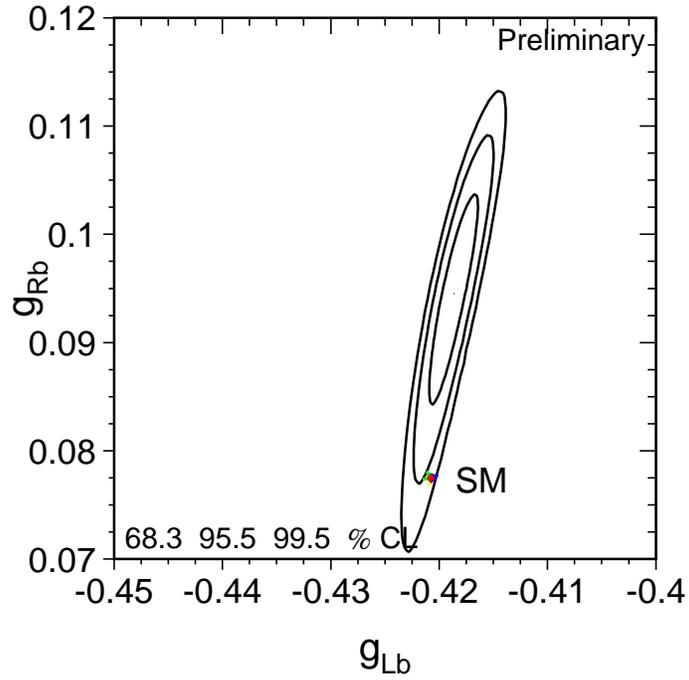,width=0.6\linewidth}
    \vskip -0.8 cm
    \caption{Determination of the left- and right-handed bottom quark
    coupling to the $Z$
    from precision measurements  \cite{lepeww}.}
    \label{fig:fig4}
  \end{center}
\end{figure}

There have been several discussions about the possibility of
this tension to be due to some new-physics effect in 
${\cal A}_b$, since, on the other hand,
 ${\cal A}_\ell(SLD)$ agrees quite well with
the leptonic forward-backward asymmetries measured 
at LEP \cite{Gambino:2003xc}. 
As seen from the fitted values of $g_{Lb}/g_{Rb}$
in Fig.~\ref{fig:fig4}, such new physics would imply an excess in $g_{Rb}$
of the order of 25\% , and no relevant effect in $g_{Lb}$,
which seems to require quite {\it ad hoc} models \cite{choud}.

More conservatively, one can think that
the ${\cal A}_{fb}^{0,b}$ is due either to some statistical fluctuation
or to some neglected systematics.
Then, the general pattern of precision measurements indicates that the SM
Lagrangian describes coherently the Z and W coupling to fermions
at the one-loop level, if a Higgs boson with mass not too far from the present
direct lower limit will be found.
\begin{figure}[hbtp]
  \begin{center}
    \epsfig{file=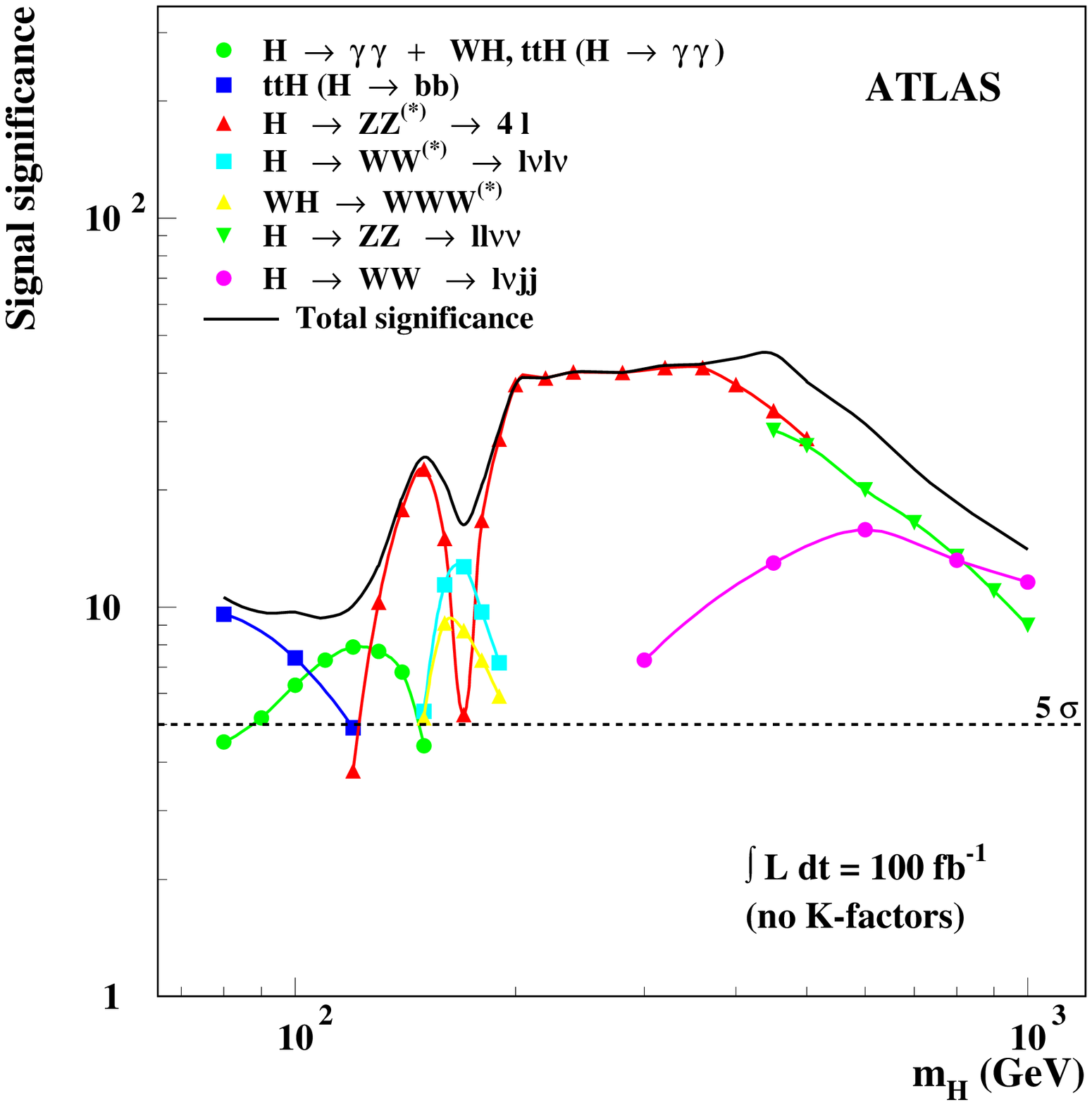,width=0.6\linewidth}
    \caption{Significance of different SM Higgs-boson 
    search modes in ATLAS for 100 fb$^{-1}$
    \cite{TDR}.}
    \label{fig:fig5}
  \end{center}
\end{figure}
Such a Higgs boson will definitely be in the realm of the LHC searches, 
as can be seen 
in Fig.~\ref{fig:fig5}, where the significance of the signal corresponding to 
the production and decay via different channels of a SM Higgs boson
is shown. If the Higgs boson is there, the LHC is going to unveil it !
\section{What Else ?}
Is there anything else we can expect to learn at the LHC ?
The answer to this question is crucially connected to another question :
do we have today in high-energy physics {\it direct} hints of the existence of
new phenomena {\it at the TeV energy scale}
that are not predicted by the SM ?

We do not, unfortunately !

On the other hand, we are pretty sure to be at the threshold of a revolution
in our knowledge of  the physics of fundamental interactions.
This is so for purely {\it conceptual} reasons, as I will try to explain in the
following.

We know today that the SM is not the {\it final} theory.
It is not, for a number of different reasons.
First of all, the minimal SM does not predict the neutrino mass, that is by 
now a well-proved experimental fact.
Second, and more dramatically,
it does not include a quantum description of the gravitational force.
We know that at energy scales near the Planck mass ($M_{Pl}\sim 10^{19}$ GeV),
where the gravitational force is expected to become as intense as 
other interactions, some radical change must enter our description of 
fundamental interactions. \\
There are then other unsolved problems in the SM, out of which I just list
the best known :
\begin{itemize}
 
\item big hierarchy in energy scales ;

\item missing coupling unification ;

\item presence of dark matter in the universe ;

\item matter-antimatter asymmetry in the universe ;

\item value of the cosmological constant ;

\item origin of the (many) fundamental parameters
       (couplings, masses, mixing angles) ;
\item \dots
\end{itemize}
While in general we do not know  which is the energy scale relevant for the
solution of most of the above issues, 
the {\it hierarchy
problem} is definitely pointing us to new phenomena at the TeV scale.
Most probably, also  the existence
of {\it dark matter} in the universe is implying new physics at the same scale.

Let us look into the {\it hierarchy problem} first.
This is connected to the instability of the scalar fields under
radiative corrections. The Higgs boson is the only
elementary scalar field entering the SM.
Contrary to other fields describing all the known particles and interactions,
that benefit from the protection of symmetry principles,
the radiative corrections to scalar fields behave rather wildly with respect
to the theory cut-off $\Lambda_{SM}$.
In particular, the corrections to the Higgs-boson mass suffer from quadratic
divergences:
 \begin{eqnarray}
\delta m_H^2&=&\frac{3G_F}{4\sqrt{2}\pi^2}\left( 2m_W^2+m_Z^2+m_H^2-4m_t^2
\right) \Lambda_{\rm SM}^2\nonumber \\
&=&-\left( 200~{\rm GeV}\frac{\Lambda_{\rm SM}}{0.7~
{\rm TeV}}\right)^2,
\label{ger}
\end{eqnarray}
where the dominant contribution comes from the large top mass.
If one wants to extend the validity of the SM
up to very large energy scales, like $M_{Pl}$ (where we know that the SM is
going to fail), or even like 
a Grand-Unification-Theory (GUT) scale $M_{GUT}\sim 
10^{16}$ GeV, in order to have a $m_H$ value consistent with experiments
(i.e., $\sim 200$ GeV), a very {\it unnatural} cancellation
between the tree-level $m_H$ value and the radiative correction
in Eq.~(\ref{ger}) must occur.
 In particular, if $\Lambda_{SM}\sim M_{Pl}$
one has to adjust the tree-level value over about 34 digits !
On the other hand,
a {\it natural} extension of the SM should imply new phenomena
at the  scale $\Lambda_{SM}\sim 1$ TeV, that quite straightforwardly 
connects to the EWSB scale. \\
A $pp$ collider of c.m. energy 14 TeV is the ideal experimental set-up
where to explore the characteristics of these new phenomena.

For the last 25 years, an intense theoretical research
has been trying to ``parametrize'' the scale $\Lambda_{SM}\sim 1$ TeV
in terms of a theoretical model that should be as consistent and predictive
as possible. \\
The main strategies adopted imply
\begin{itemize}
 
\item new symmetries ;

\item new interactions ;

\item new spacial dimensions ;

\end{itemize}
or a combination of them.

LEP results put strong constraints on the possible structure of the theoretical
model. Precision measurements powerfully limit possible contributions
to different observables coming from new physics.
If one think of the SM Lagrangian as a low-energy effective theory,
one expects new physics will contribute with non renormalizable
higher-dimension operators. The most relevant should be the
dimension-six operators ${\cal O}_{6,i}$ , with a characteristic scale
$\Lambda_{6,i}$, that enter the Lagrangian via
a $\Lambda_{6,i}^{-2}$ suppressed coupling
 \begin{equation}
  {\cal L}_{i}=\pm \frac{1}{\Lambda_{6,i}^{2}}\;{\cal O}_{6,i},
 \end{equation}
  where the positive/negative sign refers to constructive/destructive
  interference with the SM model Lagrangian.
  In Table~\ref{tab:1}, in order to be as conservative as possible,
  only ${\cal O}_{6,i}$ operators that preserve the local and global 
  SM symmetries,
  and satisfy a criterion of Minimal Flavor Violation (MFV)
  are considered \cite{Giudice:2003nc}.
  Then, one finds that in general $\Lambda_{6,i}>5-10$ TeV.
\begin{table}
\begin{center}
\caption{90\% CL limits on the scale $\Lambda_{6,i}$ (in TeV)
of dimension-six
operators ${\cal O}_{6,i}$ 
for constructive and destructive ($\pm$) interference with the
SM contribution  (from \cite{Giudice:2003nc}). 
The limits on the operators relevant to LEP1
are derived under the assumption of a light Higgs. 
\label{tab:1}} \vspace{0.2cm}
\begin{tabular}{c|c|c|c|}
&${\cal O}_{6,i}$&$\Lambda_{6,i}\;(-)$&$\Lambda_{6,i}\;(+)$\\
\hline
\raisebox{0pt}[12pt][6pt]{LEP1} &
\raisebox{0pt}[12pt][6pt]{$H^\dagger \tau^a HW^a_{\mu\nu}B^{\mu\nu}$} &
\raisebox{0pt}[12pt][6pt]{10} &
\raisebox{0pt}[12pt][6pt]{9.7} \\
\raisebox{0pt}[12pt][6pt]{} &
\raisebox{0pt}[12pt][6pt]{$| H^\dagger D_\mu H|^2$} &
\raisebox{0pt}[12pt][6pt]{5.6} &
\raisebox{0pt}[12pt][6pt]{4.6} \\
\raisebox{0pt}[12pt][6pt]{} &
\raisebox{0pt}[12pt][6pt]{$iH^\dagger D_\mu H\bar L \gamma^\mu L$} &
\raisebox{0pt}[12pt][6pt]{9.2} &
\raisebox{0pt}[12pt][6pt]{7.3} \\
\hline
\raisebox{0pt}[12pt][6pt]{LEP2} &
\raisebox{0pt}[12pt][6pt]{$\bar e \gamma_\mu e \bar\ell \gamma^\mu \ell$} &
\raisebox{0pt}[12pt][6pt]{6.1} &
\raisebox{0pt}[12pt][6pt]{4.5} \\
\raisebox{0pt}[12pt][6pt]{} &
\raisebox{0pt}[12pt][6pt]{$\bar e \gamma_\mu \gamma_5 e \bar b 
\gamma^\mu \gamma_5 b$} &
\raisebox{0pt}[12pt][6pt]{4.3} &
\raisebox{0pt}[12pt][6pt]{3.2} \\
\hline
\raisebox{0pt}[12pt][6pt]{MFV} &
\raisebox{0pt}[12pt][6pt]{$\frac{1}{2}({\bar q}_L\lambda_u\lambda_u^\dagger
\gamma_\mu q)^2$} &
\raisebox{0pt}[12pt][6pt]{6.4} &
\raisebox{0pt}[12pt][6pt]{5.0} \\
\raisebox{0pt}[12pt][6pt]{} &
\raisebox{0pt}[12pt][6pt]{$H^\dagger {\bar d}_R \lambda_d\lambda_u
\lambda_u^\dagger \sigma_{\mu \nu}q_LF^{\mu\nu}$} &
\raisebox{0pt}[12pt][6pt]{9.3} &
\raisebox{0pt}[12pt][6pt]{12.4} \\
\hline
\end{tabular}
\end{center}
\end{table}

These limits seem in contradiction with the request
of new physics at a scale as low as $\Lambda_{SM}\sim 1$ TeV. Actually, 
they can be  translated into quite severe requirements on the
model that wants to describe  new phenomena at 1 TeV.
In  particular, one can see that, in order
not to conflict with the bound $\Lambda_{6,i}>5-10$ TeV,
  new physics should be {\it weakly} coupled, 
and should not contribute very much at tree level.
On the other hand, {\it strongly} interacting models seem indeed to be
excluded by the dimension-six operator analysis up to scales of the order
of 5 to 10 TeV.
\section{Solutions to the Hierarchy Problem}
Missing any clear hint from experiments,
radically different solutions have been proposed to solve 
the hierarchy problem connected to the naturalness of the EWSB scale.

{\bf Supersymmetry} is one of the oldest \cite{susy}. 
It solves the problem of quadratic
divergences in the Higgs sector by introducing a new symmetry that connects 
different-spin particle. The particle spectrum of the SM 
is doubled, and the properties of convergence of the corresponding
field theory are drastically improved. In the limit of exact supersymmetry
(i.e., degenerate particles of same quantum numbers but  spin 
differing by 1/2),
the quadratic divergences of the scalar fields vanish.
But exact supersymmetry is not realized in nature. Then, the $m^2_H$
one-loop corrections become proportional to the splitting
in the squared masses of supersymmetric partners.
Supersymmetric partner masses of the order of (0.1$-$1) TeV would
naturally solve the EWSB scale problem.

{\bf Technicolor} has also been considered as a possible solution to the
hierarchy problem for a long time \cite{strongly}.
In this case, the Higgs boson is not an elementary scalar particle,
but a condensate of {\it technifermions} hold together by
a new (QCD-like) very strong interaction, with a scale
$\Lambda_{TC}\sim 10^3 \Lambda_{QCD}$.
Although the basic version of technicolor seems to be disfavored by LEP
precision measurements, theories with 
a very slow running of the technicolor coupling ({\it walking technicolor})
can evade the LEP constraints.

Models with {\bf large compactified extra dimensions} \cite{extra}
are quite younger than
 the previous ones, and attack the hierarchy problem from
a completely different side.
 The hierarchy between $m_H$ and $M_{Pl}$
is dissolved because the latter is not the real gravity interaction scale, that
is lowered down to the TeV scale. On the other hand,
there are new compactified spacial dimensions where the graviton also lives, 
and where most of the intensity of the ``strong" gravitational interaction
corresponding to a lower ``$M_{Pl}$"
is diluted away. In this framework, one should observe some deviation from the
Newton law at small interaction distances.

In {\bf Little Higgs models} \cite{little}, 
the SM symmetries are enlarged in such
 a way that the  Higgs boson becomes a pseudo-Goldstone boson, and its
  mass corrections vanish at one loop.
 On the other hand,
 the two-loop $m_H$ corrections allow to push $\Lambda_{SM}$ up to about 
 10 TeV without giving rise to fine-tuning problems.
 In this case the hierarchy problem would be just postponed
 by an order of magnitude in the energy scale. 
 
 In  {\bf Higgless
 extra-dimensional
 theories}, an extra dimensional component of the gauge field is
 interpreted as a Higgs field \cite{higgless}. 
 
 Although quite fascinating, 
 many of the new models seem anyhow not to fit
 the electroweak precision constraints. \\
 Still further new ideas and possibilities to solve the hierarchy problem
 could appear before the starting of the LHC.
  \section{Good and Bad Points in Supersymmetry}
At present,  the most promising way to extend the SM and solve
the hierarchy problem is to introduce supersymmetry.
This statement is based on many important facts.

First, supersymmetry is potentially
able to clear up all the difficulties related
to the TeV-scale physics :
\begin{itemize}
 
\item it stabilizes the $M_W$-versus-$M_{Pl}$ hierarchy ;

\item it explains the origin of the EWSB (by the large top-quark 
Yukawa coupling) ;

\item  it makes the measured couplings at the $M_W$ scale consistent with
a GUT model;

\item it predicts a light Higgs boson (the Higgs quartic coupling is fixed 
by gauge couplings);

\item it has a delicate impact on the electroweak precision measurements 
(virtual supersymmetric effect are suppressed by loops, so that 
$\Lambda_{6,i}\sim 4\pi \Lambda_{SM}$) ;

\item it can have a delicate impact on FCNC processes ;

\item it can naturally explain the origin of the dark matter .
\end{itemize}

Second, supersymmetry is a weakly-coupled theory.
It dramatically improves the convergence of the radiative-correction 
pattern, and is
reliably computable.  It allows accurate and consistent theoretical
predictions even at energy scales much higher than 1 TeV.
It could be extended up to scales not too far from $M_{Pl}$ ,
or even support the {\it desert} hypothesis, that very ambitiously 
implies that,
after including supersymmetric partners at the TeV scale,
no new phenomenon appears when increasing the energy up to the GUT scale.

The fact that supersymmetry is able to reconcile the SM with
a GUT model is considered today the most direct phenomenological evidence
for supersymmetry. The value of the electroweak coupling constants
$\sin^2\theta_W$ and $\alpha(M_Z)$ measured
at LEP evolve up to the GUT mass scale $M_{GUT}$ into a unified constant
$\alpha_{GUT}$ that, when evolved back to the $M_Z$ scale in the SU(3)
coupling, gives by far a too low value for the
strong coupling constant $\alpha_s(M_Z)$.
Also, the unification scale  $M_{GUT}$ would induce a too fast proton decay.
All this can be cured quite naturally by the introduction of  supersymmetry
with partner masses at the TeV scale
\cite{Langacker:1995fk}. The coupling evolution slows down, and
$\alpha_s(M_Z)$ can match its experimental value, while $M_{GUT}$
is increased up to a value consistent with the limits on the proton decay rate.

So far, everything looks perfect with supersymmetry.
But there are two bad points, one on the experimental side
and one on  the theoretical side, that have weakened up to now
the cause for supersymmetry ! After more than 20 years
of experimental searches for supersymmetric partners, no positive signal
has been detected. We just got limits on the allowed parameter space. \\
The implications of this fact are somehow made less dramatic by the
second bad point: there is a remarkable arbitrariness in the construction
of theoretical models
for the supersymmetry breaking on which the spectrum of the supersymmetric
partners is crucially based.
The only solid constraint comes from the fact that, in order to 
stabilize the hierarchy  $M_W$-versus-$M_{Pl}$, the partner masses
should be in the range $\sim (0.1\div 1)$ TeV.

Anyway, since we do not want the breaking to spoil the good 
convergence properties
of the theory, the spontaneous breaking of supersymmetry
should be susceptible to be parametrized by {\it soft} operators,
of dimension $\leq 3 \;$ \cite{Chung:2003fi}
\begin{eqnarray}
-{\cal L}_f &=&  Y_{\alpha \beta \gamma} \;\Phi^{\alpha}
\tilde \Phi^{\beta} \tilde \Phi^{\gamma} +
\mu \tilde H_u \tilde H_d + h.c. \nonumber \\
-{\cal L}_{soft} &=& m^2_{\alpha} \; \Phi^{\star \tilde \alpha} \Phi^{\alpha} +
                  M_a \; \lambda^a \lambda^a  +
		  A_{\alpha \beta \gamma} \;\Phi^{\alpha} \Phi^{\beta} \Phi^{\gamma}
		  +B \mu  H_u  H_d +h.c. \;\;\;\;\;\;\;\;\;\;\;\; \nonumber \\
\mbox{with} ~~~~~~~~~~~~ ~~~~~~
\Phi^{\alpha}&=& Q_L, u^c_L, d^c_L, L_L, e^c_L,  H_u , H_d \;. \nonumber
\end{eqnarray}

This implies introducing 108 new parameters !

In general, in any realistic supersymmetry breaking model, there are
two energy scales involved : the supersymmetry breaking scale $\sqrt F$
that is the VEV of the relevant auxiliary fields in the 
supersymmetry breaking sector, and the messenger scale ${\cal M}$
that is associated to the interactions that transmit the breaking
to the observable sector (${\cal M}\sim 30 TeV \div M_{Pl}$).
The latter give rise to soft terms for
the scalar and gaugino masses :
\begin{equation}
m^2_f \sim c_f \frac{F^2}{{\cal M}^2}, ~~~~~~~~~~
M_a \sim d_a \frac{F}{{\cal M}} \; .
\end{equation}
These mass parameters have then to be evoluted through the renormalization
group equations (RGEs) from the ${\cal M}$ scale down to the TeV scale,
the one relevant to experiments.
Experiment task, after the supersymmetry discovery, will be
to determine ${\cal M}$ and the structure of the coefficients
$c_f, d_a$.

FCNC constraints imply that squark and slepton with equal quantum numbers
are either almost degenerate in mass, or almost diagonal in the Yukawa
matrices. Different breaking schemes that verify these conditions
quite naturally have been proposed : gauge mediation (GMSB)
(with quite a low ${\cal M}$) \cite{gmed}, anomaly mediation (AMSB) 
\cite{ano}, gaugino
mediation \cite{gau}.
\begin{figure}[hbtp]
  \begin{center}
    \epsfig{file=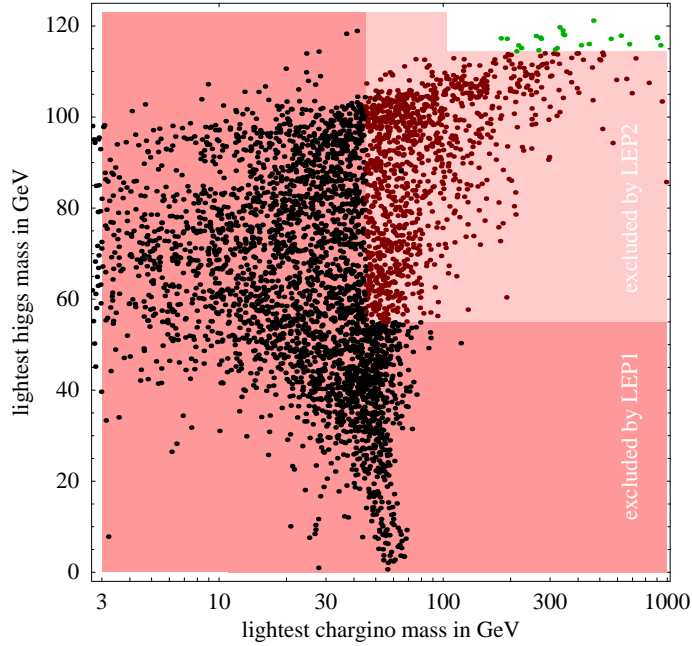,width=0.6\linewidth}
    \caption{Scatter plot of the chargino and Higgs mass obtained by
sampling the parameter space in mSUGRA. The regions
excluded by LEP1 and LEP2 searches are shown 
\cite{Giudice:2003nc,Giusti:1998gz}.}
    \label{fig:fig9}
  \end{center}
  \end{figure}
\section{Tensions in the cMSSM}
In order to maximize the predictive power 
by using as few basic parameters as possible,
most of the phenomenological studies today have been done in the 
constrained minimal supersymmetric standard model (cMSSM),
even known as minimal supergravity (mSUGRA)
\cite{Haber:1984rc}. In this framework,
a quite large supersymmetry breaking scale ($\sqrt F\sim 10^{11}$ GeV), 
is transmitted via
gravitational interactions (${\cal M}\sim M_{Pl}$) down to a mass splitting 
in the observable sector of the needed amount ($\sim 1$ TeV).
The coefficient $c_f, d_a$ are assumed to be universal, and are parametrized by
a common scalar mass $m_0$ and a common gaugino mass $M_{1/2}$ at the
scale $M_{GUT} $. Furthermore, radiative EWSB is imposed.
Indeed, in this model there are two Higgs doublets,
 and one of the Higgs scalar mass parameter, $m^2_{H_u}$,
can develop a negative value (that breaks the electroweak group
$SU(2)\times U(1)$) just as an effect of the RGEs evolution from the high scale
down to $M_W$.
From the minimization of the Higgs effective
potential one gets then the condition
\begin{equation}
\mu^2 = \frac{m_{H_d}^2 - m_{H_u}^2 \tan^2 \beta}{\tan^2 \beta -1 } - 
\frac{1}{2} M_Z^2\ .
\label{electroweak}
\end{equation} 
where
$\tan\beta= \langle H_u^0\rangle/\langle H_d^0\rangle$ 
is the ratio of the two Higgs vacuum expectation values  
(the
sign of $\mu$ remains undetermined).
For not too small $\tan\beta$ values, one has :
\begin{equation}
 M_Z^2 \sim -2\mu^2 -2 m_{H_u}^2 \ ,
\label{electroweakdue}
\end{equation} 
with all the parameter meant at low energy.
As the result of RGEs evolution,
one can  express the right-hand side of Eq.~(\ref{electroweakdue})
 as a linear combination of initial 
(high-scale) squared mass parameters, $m_0$ and $M_{1/2}$,  a common
trilinear coupling $A_0$, and $\mu$.
Then, taking into account  LEP limits on the charginos 
($m_{\chi^+_1}>103.5$ GeV \cite{lepchi})
and the light Higgs boson $h$,
the above equation implies a non-trivial cancellation between terms that are
individually quite larger that $M_Z^2$. 
A fine-tuning of the order of per cent is already implied
by the present experimental limits for the cMSSM, as  Fig.~\ref{fig:fig9}
shows \cite{Giudice:2003nc,Giusti:1998gz}.
One can easily check that the present light Higgs-mass limit pushes the squark
masses up to quite high values.
The lightest Higgs boson in the cMSSM has a tree-level mass lower than $M_Z$
that would strongly conflict with the limit
$m^{exp}_h>114$ GeV (for a quite heavy pseudoscalar) \cite{lephiggssu}.
On the other hand,  large radiative corrections are predicted at one loop
\begin{equation}
 m_h^2\; \simeq\; M_Z^2 \,\cos^2 \beta \;+ \;\frac{2\alpha_w m_{t}^4}{2\pi M_W^2
 \sin^2\beta} \;\log\frac{\tilde m^2_{t}}{m_{t}^2} \;.
\label{higgs}
\end{equation} 
In order, to be compatible with the limit $m^{exp}_h>114$ GeV, the
one-loop correction  implies both a 
large mixing in the stop
sector and a heavy stop mass, $\tilde m_t> 3.8 m_t \sim 700$ GeV.

These tensions  with experimental data in the cMSSM have motivated
theorists to consider alternative (less constrained) models, 
where the tree-level Higgs mass
is naturally larger than $M_Z$ 
\cite{Espinosa:1991gr,Espinosa:1998re,Casas:2003jx}.
\newpage
\section{Supersymmetry Signatures at the LHC}
If supersymmetry is there, it will be hard for the LHC to miss it ! $\;$
Typical experimental signatures corresponding to the production
of supersymmetric partners are in general quite striking.
The production and decay of very massive new particles
(most of the times the strongly-coupled squarks and gluinos)
give rise to large effective mass ($M_{eff}$) events 
with high jet multiplicity and high missing $p_T$, arising from the presence
of the undetected stable lightest neutralino in the particle  decay chains.
\begin{figure}[hbtp]
\vskip 1.5cm
  \begin{center}
    \epsfig{file=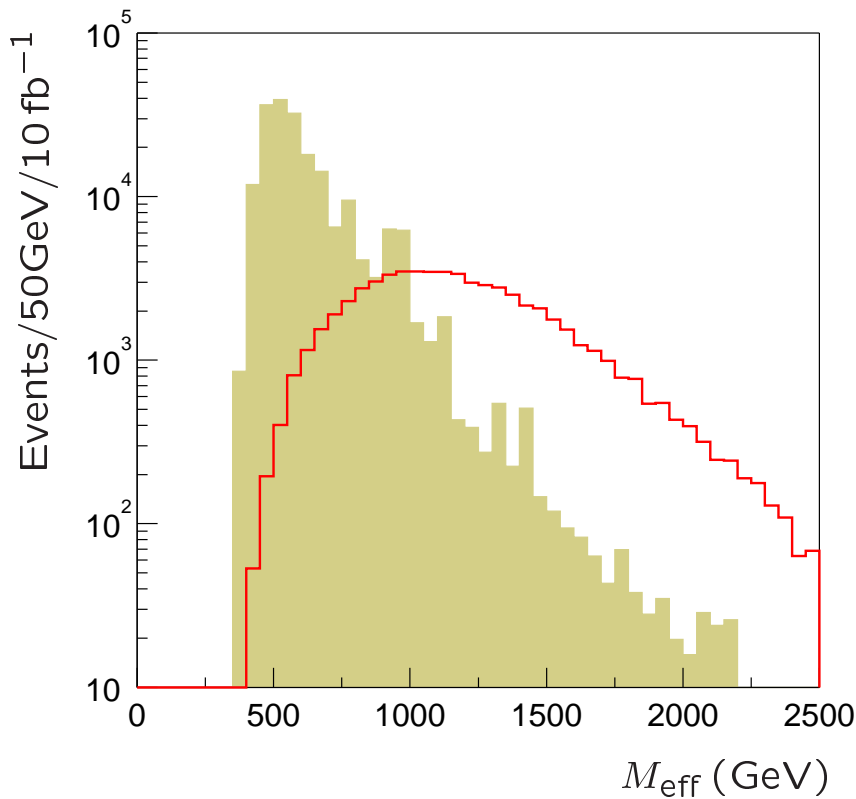,width=0.7\linewidth}
    \caption{
    $M_{eff}$ distribution for a typical mSUGRA point and SM backgrounds 
    after cuts \cite{Paige:2003sh}.}
    \label{fig:fig10}
  \end{center}
  \end{figure}
 Fig.~\ref{fig:fig10} shows 
 how the corresponding $M_{eff} 
(=E_T^{miss} + \sum_{jet} E_{T}^{jet})$
distribution in mSUGRA
compares with the SM background, for $M_{1/2}=300$ GeV,
$m_0=100$ GeV, the trilinear scalar coupling $A_0=0$, $\tan\beta=10$
and sgn $\mu=+\;$.
This is in general a very robust signature, and will be hard to miss it !

\begin{figure}[hbtp]
  \begin{center}
    \epsfig{file=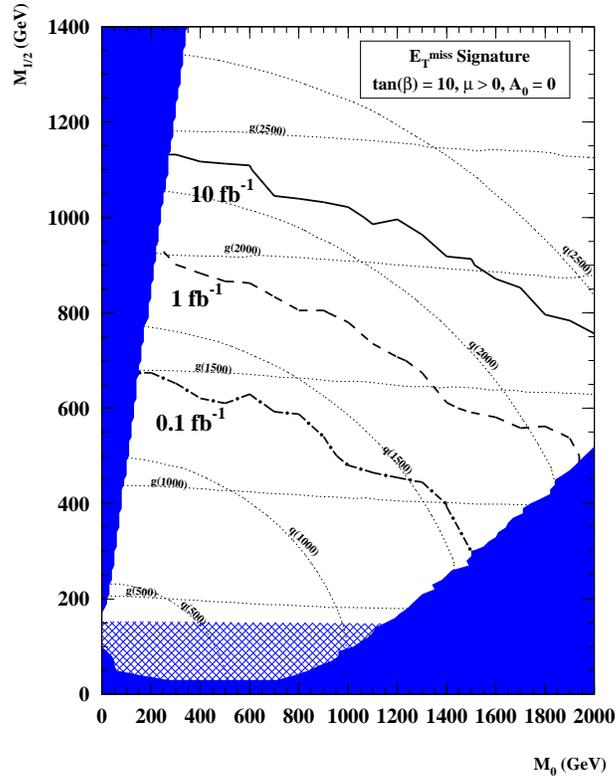,width=0.5\linewidth}
    \caption{LHC search limits for the $E_T^{miss}$ signal
    in cMSSM, with
    integrated  luminosities of 0.1, 1 and 10 $fb^{-1}$ \cite{Tovey:jc}.}
    \label{fig:fig11}
  \end{center}
  \end{figure}

In  Fig.~\ref{fig:fig11}, the $(m_0, M_{1/2})$ parameter space
covered by the $E_T^{miss}$ signature for $\tan\beta=10$, and integrated
 luminosities of 0.1, 1 and 10 $fb^{-1}$ is shown.
One month of running at ``low" luminosity ($\sim 10^{33}$ cm$^{-2}$s$^{-1}$),
that roughly corresponds to the 1 $fb^{-1}$ curve,
will be sufficient for the discovery of squark and gluinos as massive as 
1.5 TeV !

\begin{figure}[hbtp]
  \begin{center}
    \epsfig{file=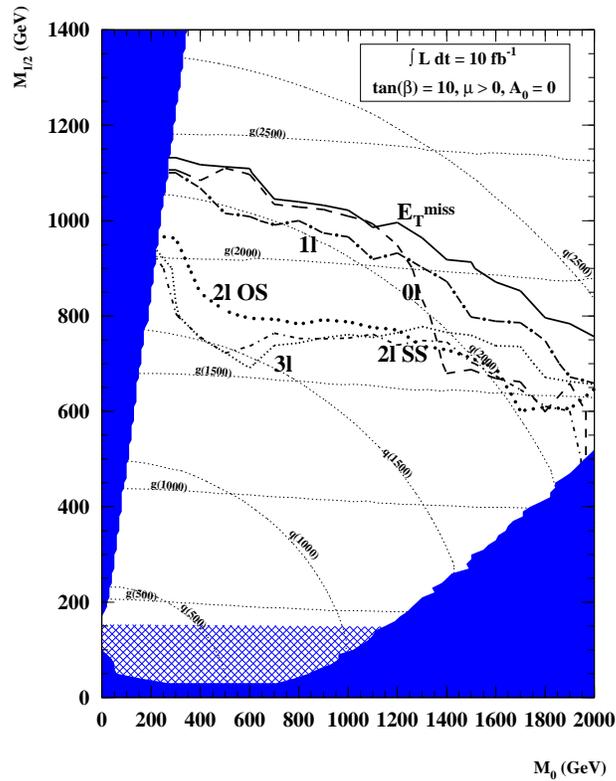,width=0.5\linewidth}
    \caption{LHC
    search limits for the cMSSM model in various 
    leptonic channels for 10 fb$^{-1}$
    \cite{Tovey:jc}.} 
    \label{fig:fig12}
  \end{center}
\end{figure}

Additional signatures have also been considered.
In  Fig.~\ref{fig:fig12}, 
the LHC search limits  based on the presence of
a number of isolated leptons  in the final state are considered.
In other supersymmetric models, like in the GMSB model,
isolated photons can be present in the final state, or even charged heavy
semi-stable particles, behaving like  heavy muons in the detectors 
\cite{Ambrosanio:2000ik}.

\begin{figure}[hbtp]
  \begin{center}
    \epsfig{file=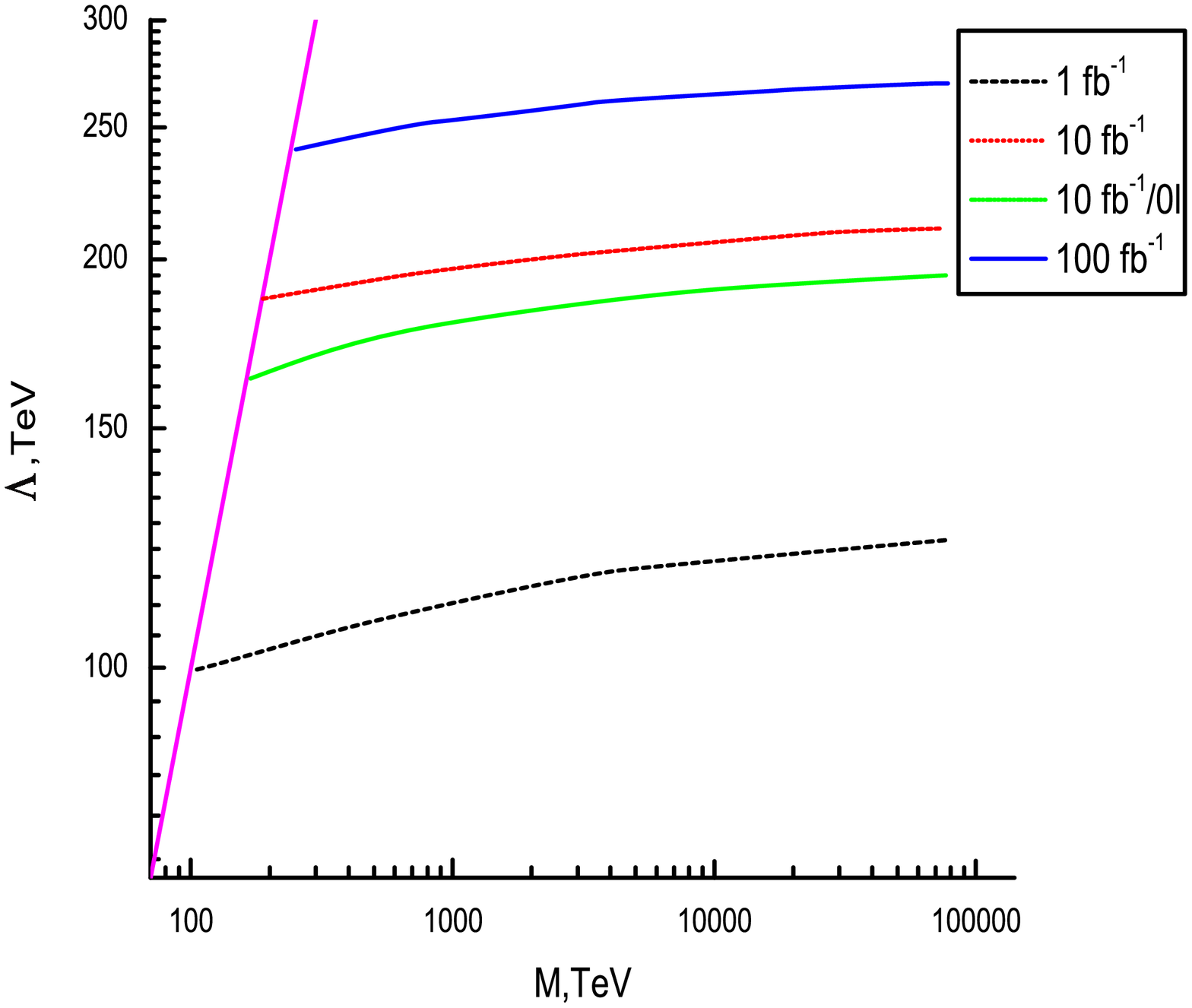,width=0.6\linewidth}
    \caption{The sensitivity reach of ATLAS to a  GMSB signal 
    \cite{Duchovni:2004yd}.}
    \label{fig:fig13}
  \end{center}
  \end{figure}

\begin{figure}[hbtp]
  \begin{center}
    \epsfig{file=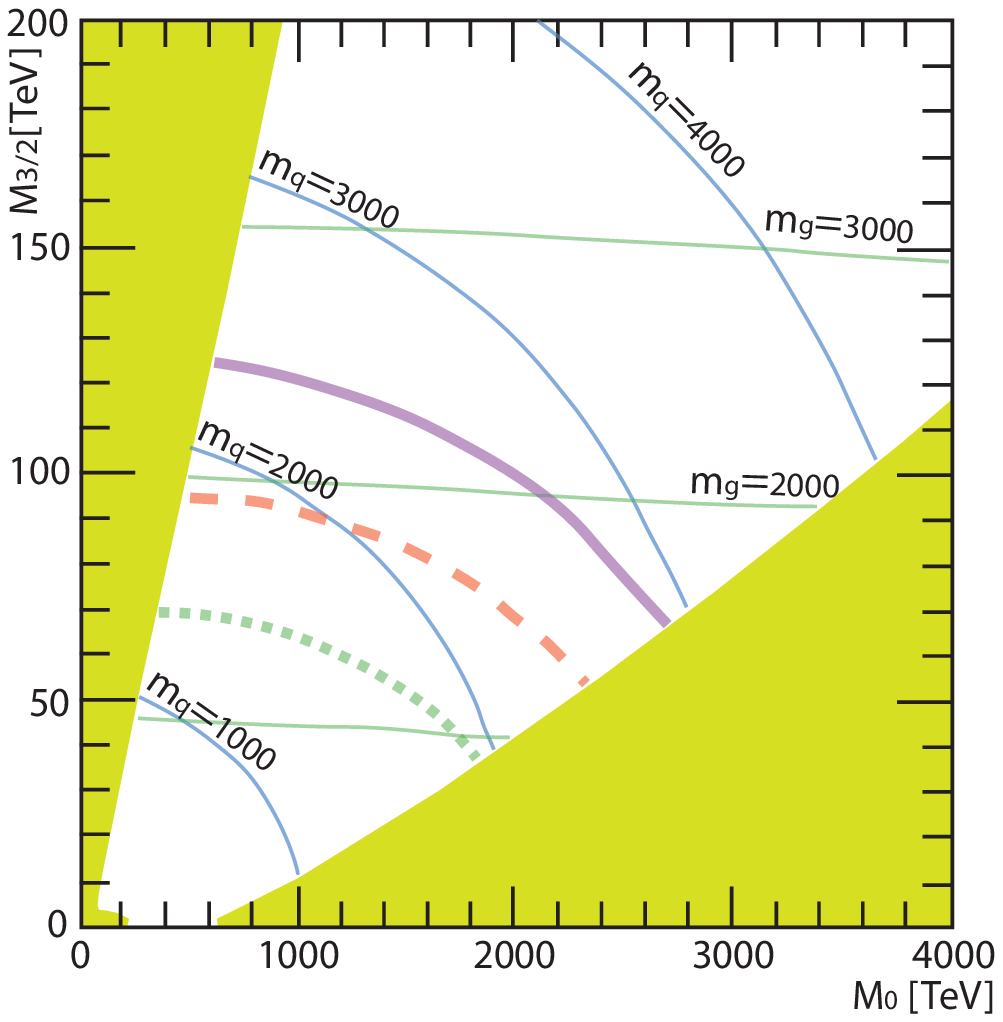,width=0.6\linewidth}
    \caption{The sensitivity reach of ATLAS in the AMSB parameter space
  for luminosities of 1 (short-dashed), 10 (long-dashed)
  and 100 (solid) $fb^{-1}$ \cite{Duchovni:2004yd}.}
    \label{fig:fig14}
  \end{center}
  \end{figure}

In  Fig.~\ref{fig:fig13}, the expected reach in the GMSB model parameter 
$\Lambda$ for different integrated luminosities is shown \cite{Duchovni:2004yd}.
In  Fig.~\ref{fig:fig14}, the expected reach for integrated
luminosities of 1, 10 and 100 $fb^{-1}$ is presented 
versus the AMSB model parameters  \cite{Duchovni:2004yd}.
  \\
  The coverage of the parameter space that is relevant 
  to the  solution of the hierarchy problem is in general guaranteed 
  at the LHC for any supersymmetry-breaking model considered !
 \newpage
  \section{Dark Matter and LHC Searches}
\begin{figure}[hbtp]
 \vskip 1 cm
  \begin{center}
    \epsfig{file=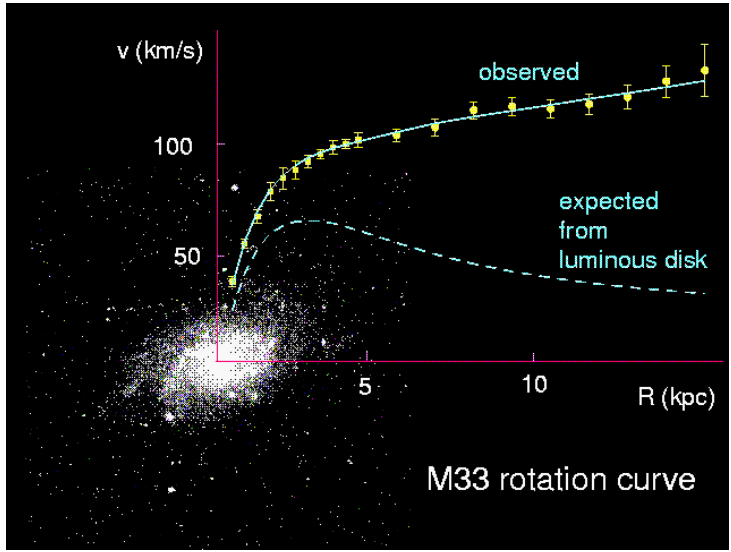,width=0.6\linewidth}
    \caption{The observed rotation curve of the dwarf spiral galaxy M33 extends
    considerably beyond its optical image (shown superimposed), from
  \cite{roy}.}
    \label{fig:fig15}
  \end{center}
  \end{figure}

  The need for {\it dark matter} is a long-standing issue in modern cosmology
  \cite{Sahni:2004ai}. 
  A number of astrophysical and cosmological observations indicate
  that a substantial fraction (perhaps as much as 30\%) of the energy density 
  in the Universe is
  due to non relativistic, non baryonic, non luminous matter.
  The presence of an ``unseen" component of matter in individual galaxies
  has been well established by looking at the rotation curve 
 $v(r)$ of galaxies (on a sample of over one thousand spiral galaxies).
  The Kepler's third law predicts
  \begin{equation}
v(r) = \sqrt{\frac{G M(r)}{r}} ,
\label{kepler}
\end{equation} 
where $r$ is the radial distance from the galactic center.
As can be seen in  Fig.~\ref{fig:fig15}, the observed curve, instead of 
behaving
as $1/r^{1/2}$, flattens out 
at large distances from the luminous disk. This implies a mass increasing
like $M(r)\propto r$ in the region where no luminous mass can account for such
an increase.
In particular, in order to explain the galactic rotation curves, one
needs about one order of magnitude more matter in the galaxies than the one corresponding
to the ``seen" component.
It is not easy to explain rotation curves via possible modifications of the 
Newton law. Altogether, this evidence is quite tight and solid,
not depending on particular cosmological assumptions.

The microscopic nature of dark matter is presently not known.
In particle physics, one can quite naturally imagine that it is
made of {\it thermal relics}. A thermal relic is a stable neutral 
particle that was in thermal equilibrium at the starting universe, and, due to
the universe expansion, at a certain time ({\it freeze-out} time)
decouples.
The present {\it relic density}
of these particles, $\Omega_{DM}$, depends on the moment
they decouple, that, in turn, depends on the time 
when the particle annihilation
rate equals the expansion rate of the universe. Then, one has
  \begin{equation}
  \Omega_{DM} \propto \frac{1}{\langle v_a \; \sigma_{ann}\rangle},
\label{dmatter}
\end{equation} 
where $v_a$ is the relative velocity of two annihilating particles and 
the constant of proportionality is computable.
By using the recent accurate WMAP measurement $\Omega_{DM} h^2=0.113\pm 0.009$
(where $h=H_0/100km/s/Mpc = 0.7$ is the current Hubble expansion rate)
\cite{spergel03}, one finds
  \begin{equation}
\langle v_a \; \sigma_{ann} \rangle \sim 1 \; {\rm pb}\;.
\label{sigann}
\end{equation} 
This rate is the one characteristic of electroweakly interacting particles
with mass $m$ of the order of 100 GeV, that is 
$\sigma_{ann} \sim \alpha^2/m^2$ !
As far as we know today, it could have been different from this
value by orders of magnitude. \\
All this leads quite naturally to the {\it weakly interacting massive particle}
(WIMP) hypothesis : dark matter is made of stable neutral particles 
as heavy as $(100-1000)$ GeV, weakly coupled to ordinary matter.
This implies that both production cross sections and signatures of this new kind
of matter are of interest for the  LHC ! High $p_T$ jets with large missing 
transverse energy corresponding to WIMP's production
would naturally emerge on the SM background !

\begin{figure}[hbtp]
  \begin{center}
   \vskip 1 cm
    \epsfig{file=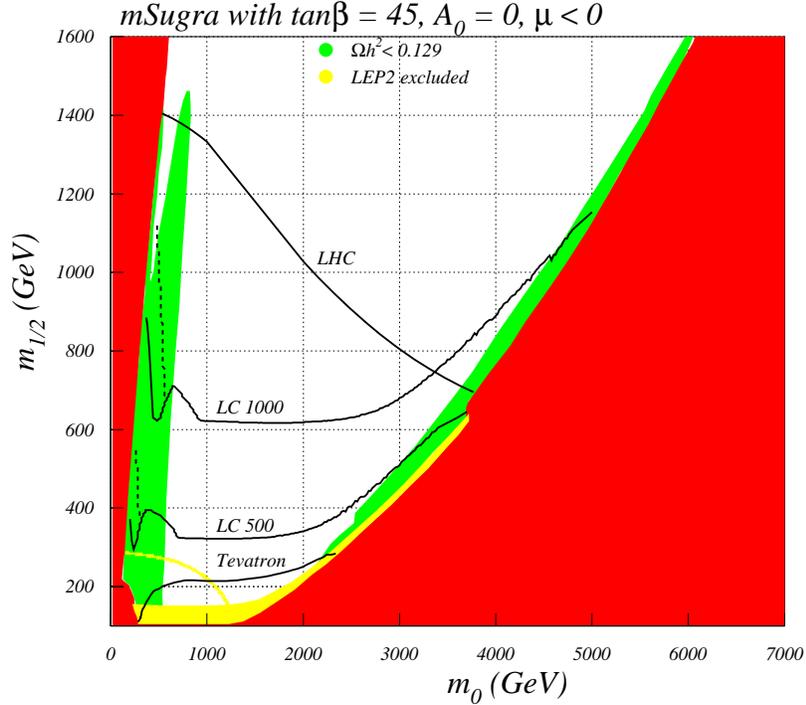,width=0.7\linewidth}
    \caption{Reach of the LHC and a linear collider (LC)
     with  $\sqrt{s}=0.5$ and 1~TeV,  
 compared with the (green) region compatible with a neutralino hypothesis
 for the WIMP's in the cMSSM, from \cite{Baer:2003ru}. }   
    \label{fig:fig16}
  \end{center}
  \end{figure}

There is a number of different theoretical models for the EWSB that
can predict  WIMP's. The oldest is supersymmetry, that, for conserved 
R parity, predicts a stable lightest neutralino very  well compatible
with a WIMP interpretation.
Fig.~\ref{fig:fig16}, from \cite{Baer:2003ru}, 
compares the reach of LHC and future 
$e^+e^-$ linear colliders in the cMSSM parameter space with the region 
that is compatible with the WMAP relic-density determination.

More recently, different theoretical frameworks
giving rise to  WIMP-like particles, like 
Universal-Extra-Dimension and Little-Higgs models,
have been considered \cite{Matchev:2004pr}.
In general, the requirements to be fulfilled by the new model are:
existence of new particles; new symmetry that makes one of the 
new particles stable; possibility to adjust the model parameters 
to make the lightest new stable particle neutral, and with the proper 
thermal relic density. 

Many possibilities can be considered,
and the LHC will be extremely helpful in discriminating among them !
\newpage

\section{Conclusions}
The EWSB sector of the SM is not yet completely tested. 
The Higgs boson is still 
to be found, and, eventually, its properties are to be tested.
In order to stabilize the EWSB scale, new physics at 
the TeV scale is required. This holds in any of the following cases ;
\begin{itemize}
 
\item the Higgs boson is light ($< 200$ GeV)
 
\item the Higgs boson is heavy ($> 200$ GeV)

\item the Higgs boson will not be found.

\end{itemize}

There are presently quite a few theoretical models that try to solve
(or at least to postpone) the $M_W$-versus-$M_{Pl}$ hierarchy
problem. Unfortunately, for no one of them  there is today any direct 
experimental evidence.
The LHC will be able to discriminate among them.

Given our confidence in the need for new physics at the TeV scale
on the one hand, and, on the other hand, the fact that up to now 
no model is supported experimentally, 
the {\it discovery} potential of the LHC is enormous, if  {\it discovery}
stands for finding {\it unforeseen} phenomena.
We are likely to be at the edge of a revolutionary time for the
physics of fundamental interactions !


\end{document}